\documentclass[12pt,peerreview,onecolumn]{IEEEtran}

\IEEEoverridecommandlockouts

\usepackage{color}
\usepackage{graphicx}
\usepackage{amsmath}
\usepackage{amssymb}
\usepackage{algorithm}
\usepackage{algorithmic}
\usepackage{amsmath}
\usepackage{multirow}
\usepackage{booktabs}
\usepackage{array}
\usepackage{amsthm}
\usepackage{stfloats}
\usepackage{caption}
\usepackage{bm}
\usepackage{subfigure}

\newcommand{\be}{\begin{equation}}
\newcommand{\ee}{\end{equation}}
\newcommand{\bea}{\begin{eqnarray}}
\newcommand{\eea}{\end{eqnarray}}
\newcommand{\ba}{\begin{array}}
\newcommand{\ea}{\end{array}}

\makeatletter

\newcommand{\Rmnum}[1]{\expandafter\@slowromancap\romannumeral #1@}
\makeatother

\newcommand{\RNum}[1]{\uppercase\expandafter{\romannumeral #1\relax}}

\title{Intelligent Reflecting Surface Assisted Multi-cell Multi-band Wireless Networks \thanks{W. Cai, R. Liu, Y. Liu, and M. Li are with the School of Information and Communication Engineering, Dalian University of Technology, Dalian 116024, China, (e-mail: wenhaocai@mail.dlut.edu.cn, liurang@mail.dlut.edu.cn, yangliu613@dlut.edu.cn, mli@dlut.edu.cn).}
\thanks{Q. Liu is with the School of Computer Science and Technology, Dalian University of Technology, Dalian 116024, China (e-mail: qianliu@dlut.edu.cn).}}

\author{Wenhao Cai,
        Rang Liu,~\IEEEmembership{Student Member,~IEEE,}
        Yang Liu,~\IEEEmembership{Member,~IEEE,}
        Ming Li,~\IEEEmembership{Senior Member,~IEEE,}
        and Qian Liu,~\IEEEmembership{Member,~IEEE}}

\begin{document}
\maketitle
\thispagestyle{empty}
\vspace{-1 cm} 
\begin{abstract}
Intelligent reflecting surface (IRS) is deemed as a promising and revolutionizing technology for future wireless communication systems owing to its capability to intelligently change the propagation environment and introduce a new dimension into wireless communication optimization.
Most existing studies on IRS are based on an ideal reflection model.
However, it is difficult to implement an IRS which can simultaneously realize any adjustable phase shift for the signals with different frequencies.
Therefore, the practical phase shift model, which can describe the difference of IRS phase shift responses for the signals with different frequencies, should be utilized in the IRS optimization for wideband and multi-band systems.
In this paper, we consider an IRS-assisted multi-cell multi-band system, in which different base stations (BSs) operate at different frequency bands. We aim to jointly design the transmit beamforming of BSs and the reflection beamforming of the IRS to minimize the total transmit power subject to signal to interference-plus-noise ratio (SINR) constraints of individual user and the practical IRS reflection model.
With the aid of the practical phase shift model, the influence between the signals with different frequencies is taken into account during the design of IRS.
Simulation results illustrate the importance of considering the practical communication scenario on the IRS designs and validate the effectiveness of our proposed algorithm.
\end{abstract}

\begin{IEEEkeywords}
Intelligent reflecting surface, practical model, multi-band system, beamforming optimization.
\end{IEEEkeywords}

\maketitle
\vspace{-0.3 cm}
\section{Introduction}
With the rapid development of wireless services and the popularizing of intelligent devices, the demand for wireless network services is growing exponentially.
This motivates research on the key technologies of the fifth-generation (5G) and beyond networks. As a promising and revolutionizing technology for future wireless communication systems, intelligent reflecting surface (IRS) can extend the coverage of wireless communications and prominently enhance the transmission quality by intelligently changing the propagation environment \cite{IEEE 1}-\cite{Online 3}. Thus, the IRS has drawn significant attention within both the industry and academic communities.

The IRS consists of a large number of nearly passive elements with ultra-low power consumption. Particularly, each element of the IRS is composed of configurable electromagnetic (EM) internals, which are capable of controlling the phase shift and amplitude of the incident EM wave in a programmable manner. Therefore, the IRS is envisioned to revolutionize the current communication optimization paradigm by integrating the smart radio environment and expected to play an important role in future wireless communications. Thus, the applications of IRS in different wireless communication scenarios have been extensively investigated with different performance metrics \cite{IRS_Liuxing}-\cite{NEW}.

In the aforementioned studies, it is assumed that each IRS element has an ideal but impractical reflection model, in which phase shift can be adjusted to any value for any frequencies.
The design of IRS with such a simple assumption can be easily implemented using classical optimization tools. However, it is extremely difficult to implement an IRS having such an ideal reflection model due to the hardware circuit limitation.
In \cite{Practical_phase}-\cite{IRS_Hardware1}, it has been illustrated that the IRS actually produces different phase shifts for the signals with different frequencies.  Particularly, the authors in \cite{Practical_phase} proposed a practical model of reflection coefficient for an IRS-aided wideband multiuser MISO system and developed a heuristic algorithm to prove that the use of the practical model can significantly improve the system performance.
However, the IRS designs with a practical model for more complex systems have not been investigated. For example, in the multi-cell multi-band system, when the IRS provides the desired phase shift to the signal of one base station (BS), it will also affect the signals of other BSs. This phenomenon will cause difficulties on the IRS design. Thus, considering the practical IRS reflection model in a realistic communication scenario is very crucial to provide more accurate guidance for the IRS beamforming optimization and other designs.
To the best of our knowledge, the IRS designs for multi-cell multi-band communication systems with a practical IRS phase shift model have not been investigated in the literature yet.

In this work, we consider an IRS-assisted multi-cell multi-band system, in which BSs belong to the same service provider but operate at different frequency bands.
We first analyze the IRS response for the signals at different frequency bands and  provide a practical frequency-dependent IRS response model to facilitate the IRS design in the multi-cell multi-band scenario.
Then, we aim to jointly design the transmit beamforming of multiple BSs and the reflection beamforming of the IRS to minimize the total transmit power subject to the signal to interference-plus-noise ratio (SINR) constrains of individual users and the practical IRS reflection model.
An iterative algorithm is introduced to sub-optimally solve this non-convex problem. Simulation results illustrate that significant performance improvement can be achieved by considering the practical model in the design of IRS for this multi-cell multi-band communication scenario.


\section{Practical IRS and System Model}
\subsection{Practical IRS Phase Shift Model}
\begin{figure}[t]
\centering
  \includegraphics[width= 3 in]{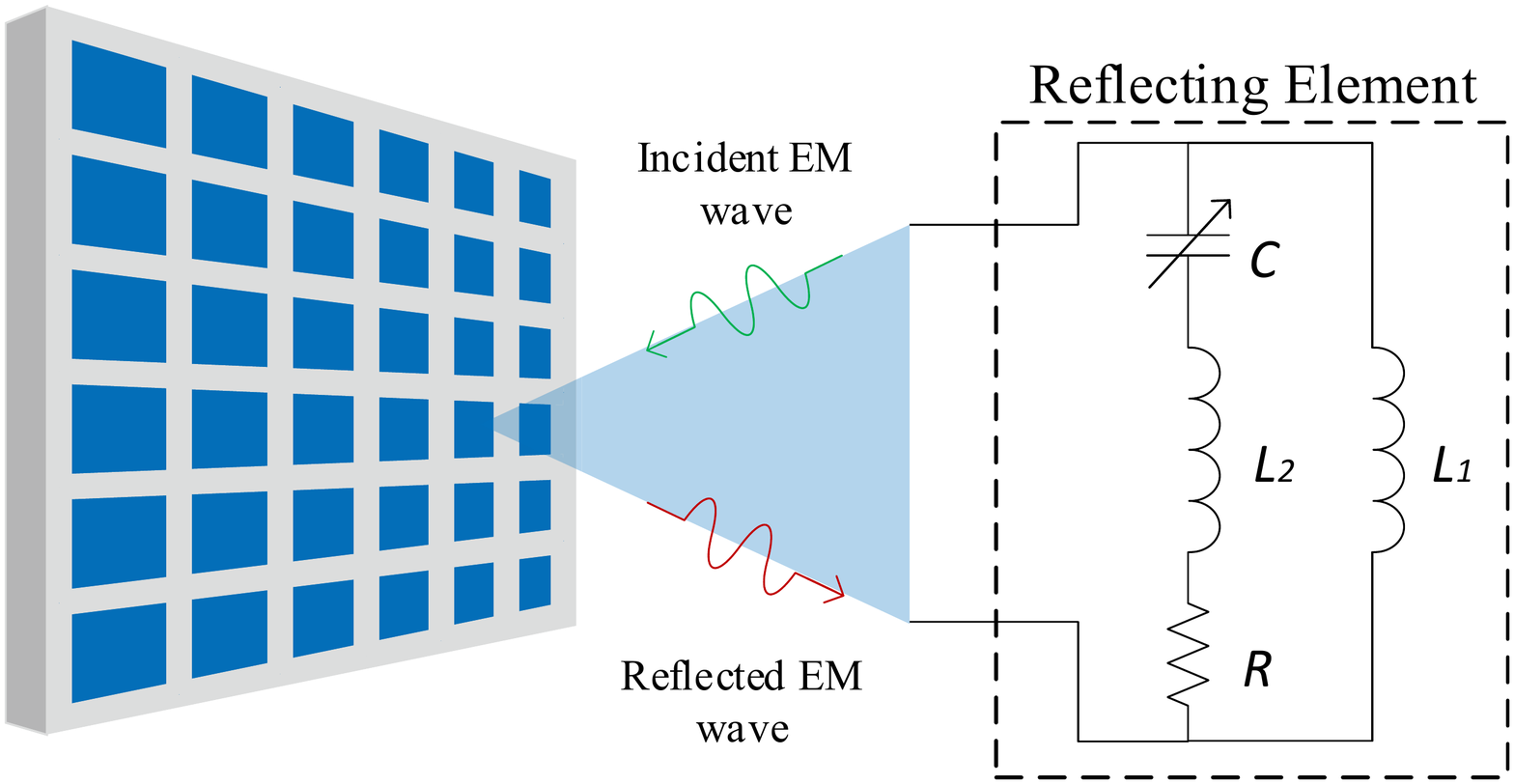}
  \caption{The equivalent circuit of a practical IRS element.}\label{fig:electric_line}
\end{figure}

IRS is a kind of reconfigurable impedance metasurface, which is composed of plenty of configurable electromagnetic (EM) internals and can reflect the incident EM wave with a desired phase shift by adjusting electrical parameters, thus collaboratively generating reflection beamforming.
Previous works \cite{Practical_phase}, \cite{Practical_phase2} have illustrated that the phase shift of the reflected signal is associated with the frequency and parameters of the reflecting circuit.
To be specific, the electrical characteristics of an IRS element are illustrated by a parallel resonant circuit, as shown in Fig. \ref{fig:electric_line}. The amplitude and phase shift of the reflected signal are controlled by selecting an appropriate capacitance $C$. For a signal of frequency $f$, the impedance of the reflecting element circuit can be written by
\begin{equation}
Z(C,f) = \frac{j2{\pi}f L_1(j2{\pi}f L_2+\frac{1}{j2{\pi}f C}+R)}{j2{\pi}fL_1+(j2{\pi}fL_2+\frac{1}{j2{\pi}f C}+R)},\label{eq:gammaz}
\end{equation}
where $L_1$, $L_2$, $C$, and $R$ denote the equivalent inductances, variable capacitance, and the loss resistance in the equivalent circuit, respectively. Then, the reflection coefficient of the $m$-th IRS element can be obtained by
\begin{equation}
\phi_m(C, f) = \frac{{Z}_m({C},f)-{Z}_{0}}{{Z}_m({C},f)+{Z}_{0}},\label{eq:gamma}
\end{equation}
where $Z_0 = 377\ \Omega$ denotes the free space impedance and $Z_m(C,f)$ denotes the $m$-th element parallel resonant circuit impedance, which can be easily obtained from the popular parallel impedance formula.

\begin{figure}[t]
\centering
  \includegraphics[width= 4 in]{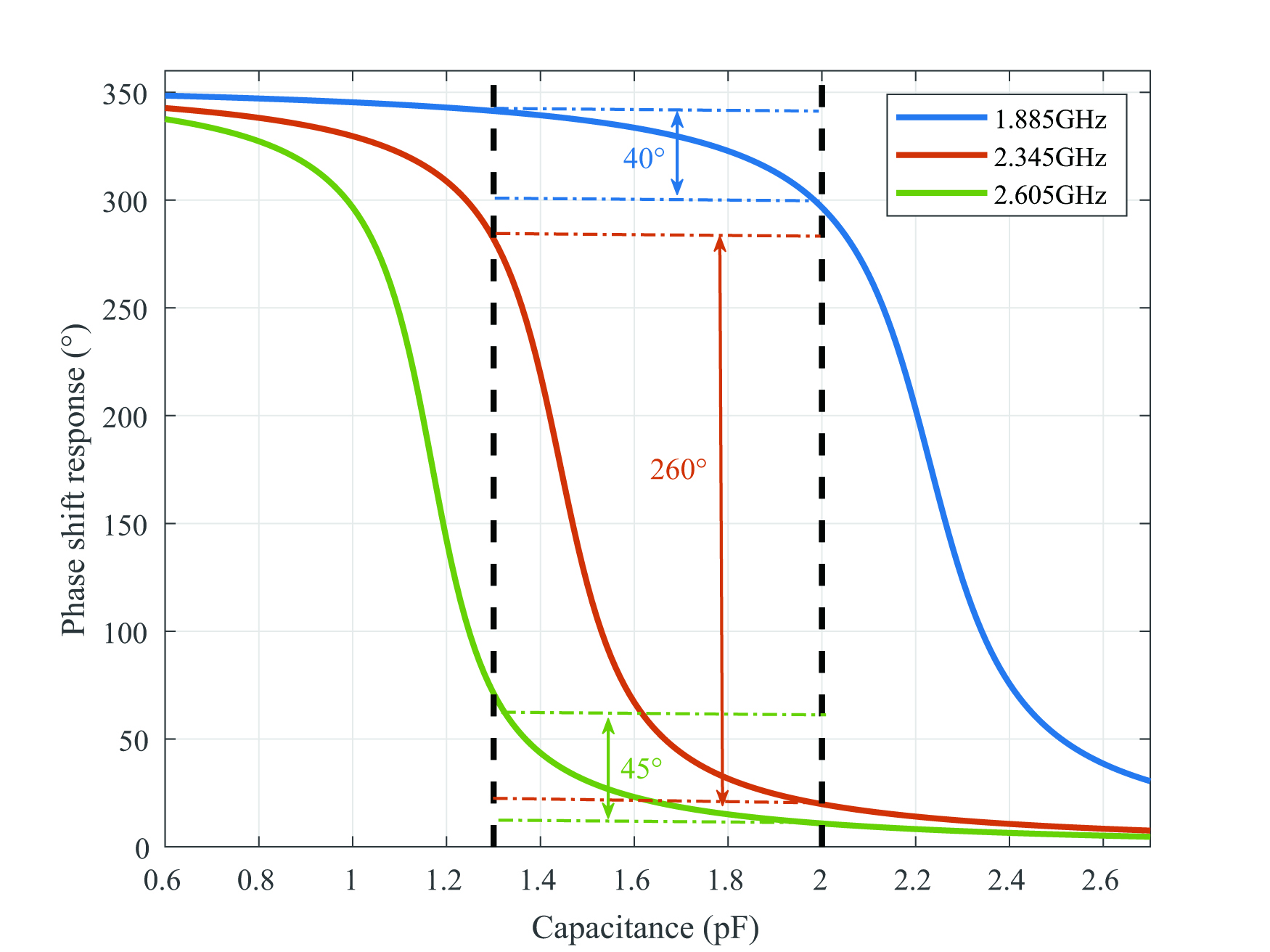}
  \vspace{-0.1 cm}
  \caption{The phase shift versus capacitance with different frequencies.}\label{fig:Mode}
\end{figure}

From (\ref{eq:gamma}), we can know when the incident signals have different frequencies, the IRS will provide different phase shifts depending on the range of capacitance in the equivalent circuit.
To illustrate this important phenomenon, we select a practical surface-mount diode SMV1231-079 to implement the IRS \cite{IRS_Realize} with practical parameters $L_1=2.5\ \mathrm{nH}, L_2=0.7\ \mathrm{nH}$, and $R=1\ \Omega$.
As an example, we consider an IRS-assisted wireless network with three BSs, which operate at $1.885\ \mathrm{GHz}$, $2.345\ \mathrm{GHz}$, and $2.605\ \mathrm{GHz}$ frequencies, respectively.
Based on equations (\ref{eq:gammaz}) and (\ref{eq:gamma}) with the above parameters, we can obtain the relationship between phase shifts and capacitances for the signals with different frequencies as shown in Fig. \ref{fig:Mode}.
Therefore, the signal of each BS has different IRS reflection coefficients.

\begin{table}[t]
\normalsize
\centering
    \caption{Practical IRS phase shifts under different range of capacitance in the communication system with three frequency bands.}
    \begin{tabular}{ccccccc}
    \hline
      & $C_1$  & $C_2$ & $C_3$ & $C_\mathrm{no}$\\
    \hline
    1.885 $\mathrm{GHz}$ & $0$  & $0$  & $(0, 2\pi]$  & $0 / 2\pi$ \\
    \hline
    2.345 $\mathrm{GHz}$ & $0$ & $(0, 2\pi]$ & $2\pi$   & $0 / 2\pi$\\
    \hline
    2.605 $\mathrm{GHz}$ & $(0, 2\pi]$ & $2\pi$  & $2\pi$  &$0 / 2\pi$\\
    \hline\label{table:1}  \vspace{-1 cm}
    \end{tabular}
\end{table}

It can be observed from Fig. \ref{fig:Mode} that, when the IRS provides a tunable phase shift for the signal with a certain frequency, the signals at other frequency bands exhibit almost fixed phase shifts.
This is because that the IRS reflection coefficient is associated with signal frequency and capacitance as shown in (\ref{eq:gamma}). Therefore, for a practical IRS element with an ascertained capacitance, it inevitably offers different phase shifts for signals with different frequencies.
Similar conclusions can also be drawn from existing researches \cite{IRS_Hardware1}, \cite{IRS_Hardware2} and \cite{IRS_Hardware3}.
For example, when the certain IRS element primarily serves a BS with the frequency of 2.345 $\mathrm{GHz}$, i.e., the capacitance of this IRS element changes from 1.3 $\mathrm{pF}$ to 2 $\mathrm{pF}$, the signal with 2.345 $\mathrm{GHz}$ carrier frequency has up to 260-degrees phase range, while the signals with 1.885 $\mathrm{GHz}$ and 2.605 $\mathrm{GHz}$ carrier frequencies have only 30-degrees and 50-degrees phase ranges.
Thus, we can merely consider that the phase shifts of the signal from the secondary BS are fixed. Moreover, since IRS commonly uses low-resolution (e.g., 1 bit or 2 bit) phase shift, the phase range for the signal from the secondary BS is less than the quantization interval, which can be simply considered unchangeable.

Therefore, we can summarize the phase shift relationship between signals with different frequencies shown in Fig. \ref{fig:Mode} in Table \RNum{1}.
Each column represents a status which indicates that the IRS primarily serves one BS (i.e., $C_1$, $C_2$, and $C_3$) or no BS (i.e., $C_\mathrm{no}$).
To illustrate this table, we can take an instance when the capacitance of a certain IRS element ranges from 2 $\mathrm{pF}$ to 2.6 $\mathrm{pF}$, which is called status $C_3$, this IRS primary serves the BS with 1.885 $\mathrm{GHz}$, i.e., the signal with this frequency has a wide tunable phase shift range within $[0, 2\pi]$ while the signals with other frequencies has almost constant phase shift around $2\pi$.
Thus, $C_s$, $s=1,2,3,$ represents the capacitance range status of the IRS which makes it provide tunable phase shift to the $s$-th BS.
Besides, $C_\mathrm{no}$ represents the capacitance range status of the IRS which makes it cannot provide any tunable phase shift for all users, just as the capacitance range from 0.6 $\mathrm{pF}$ to 0.8 $\mathrm{pF}$ in Fig. \ref{fig:Mode}. In summary, IRS is widely recognized as a frequency selective device and the property as shown in Table \RNum{1} should be considered in the IRS optimizations.

\subsection{System Model}
\begin{figure}[t]
\centering
  \includegraphics[height=2.2 in]{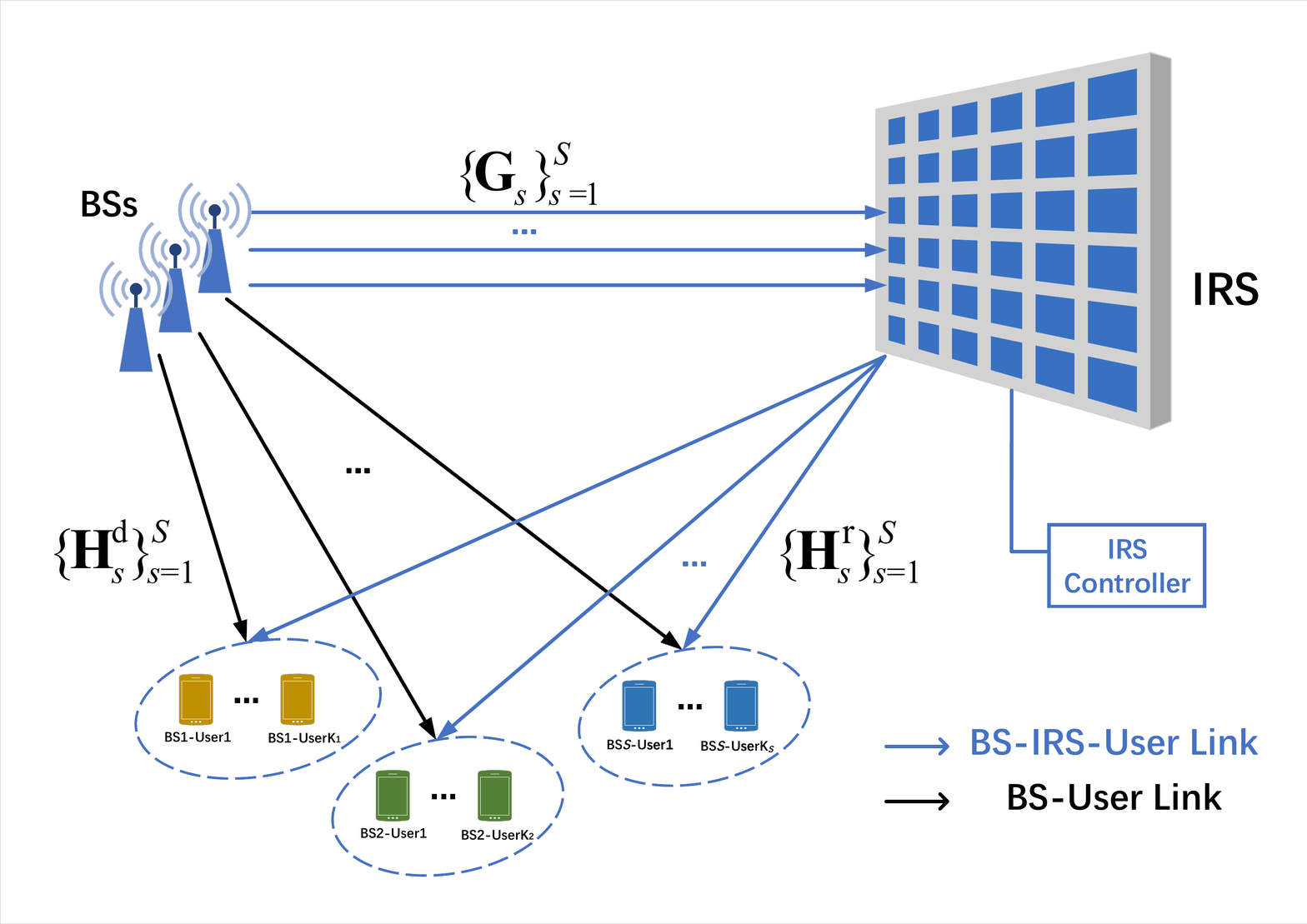}
  \caption{An IRS-aided multi-cell multi-band communication system.}\label{fig:System model}
  \vspace{-0.3 cm}
\end{figure}

In this paper, we consider a wireless communication system with $S$ BSs operating at different frequencies bands, where an IRS composed of $M$ reflecting elements is utilized to assist the communication from the BSs with $N_\mathrm{t}$ antennas\footnote{In this paper, the transmit antennas for different BS is assumed to be same for the practical deployment.} to $K_s$ single-antenna users, $s = 1, \ldots, S,$ as shown in Fig. \ref{fig:System model}.
Let $\mathcal{S} \triangleq \{1, \ldots, S\}$ denote the set of BSs with different frequency bands, which is sorted in an order from low-frequency BS to high-frequency BS.
Let $\mathcal{N}_\mathrm{t} \triangleq \{1, \ldots, N_\mathrm{t}\}$ denote the set of transmit antennas for each BS,
$\mathcal{K}_s \triangleq \{1, \ldots, K_s\}$ denote the set of the users served by the $s$-th BS,
and $\mathcal{M} \triangleq \{1, \ldots, M\}$ denote the elements of the IRS. Furthermore, we assume that the IRS is controlled by an IRS controller through an independent control link and all the channel state information (CSI) is known perfectly and instantaneously to the BSs, which can be obtained via the efficient channel estimation approaches proposed by the existing mature works.
Next, we will describe the communication procedure in details.

In this paper, we consider linear transmit precoding at the BSs where each user is assigned with one dedicated beamforming vector. Let $\mathbf{z}_{s} \triangleq [z_{1,s}, \ldots, z_{K_s,s}]^T \in \mathbb{C}^{K_s},\forall s \in \mathcal{S}$ be the transmit symbols for all users served by the $s$-th BS with $\mathbb{E}\{\mathbf{z}_{s} \mathbf{z}_{s}^*\} = \mathbf{I}_{K_s}, \forall s \in \mathcal{S}$.
The vector $\mathbf{z}_{s}$ is precoded by a precoder matrix $\mathbf{W}_s \triangleq [\mathbf{w}_{1,s}, \ldots, \mathbf{w}_{{K_s},s}],$ where $\mathbf{w}_{{k_s},s} \triangleq [w_1, \ldots, w_{N_\mathrm{t}}]^T \in \mathbb{C}^{N_\mathrm{t }\times 1}, \forall s \in \mathcal{S}$.
Hence, the complex baseband transmitted signal at the $s$-th BS can be expressed as
\begin{equation}
   \mathbf{x}_s =  \mathbf{W}_{s}\mathbf{z}_{s}, \forall s.
\end{equation}

In the considered IRS-assisted multi-cell multi-band system, the baseband equivalent channels from the $s$-th BS to IRS, from the IRS to the user $k_s$, and from the $s$-th BS to the user $k_s$ are denoted by $\mathbf{G}_{s} \in \mathbb{C}^{M \times N_\mathrm{t}}$,
$\mathbf{h}_{\mathrm{r},k_s,s} \in \mathbb{C}^{M}$ and
$\mathbf{h}_{\mathrm{d},k_s,s} \in \mathbb{C}^{N_\mathrm{t}}$,
respectively, $\forall k_s \in \mathcal{K}_s, \forall s \in \mathcal{S}$.

According to the practical IRS phase shift model, when the IRS provides the desired phase shift to the signal from a certain BS, due to the frequency-selective characteristic of the IRS, it will affect the signals from other BSs in a different manner. This phenomenon will cause difficulty to the IRS phase shift design. To facilitate the description of the system model and illustrate how the IRS affect signals from different BSs, we define an IRS indicator matrix as
\begin{equation}
   \mathbf{A} \triangleq [\mathbf{a}_1, \ldots, \mathbf{a}_M] \triangleq
   \left[\begin{array}{c}
\underbrace{a_{11}, a_{12},   \ldots,  a_{1M}} \\
\tilde{\mathbf{a}}^T_1\\
\vdots \\
\underbrace{a_{S1},  a_{S2},   \ldots, a_{SM}} \\
\tilde{\mathbf{a}}^T_S\\
   \end{array}\right],
   \label{eq:i}
\end{equation}
where $a_{s,m}$ is a binary decision variable, i.e., $a_{1,m} \in \{0,1\}, \forall s \in \mathcal{S}, \forall m \in \mathcal{M}$. To be specific, $a_{s,m} = 1$ represents that the $m$-th element provides the desired phase shift to the $s$-th BS and $a_{s,m} = 0$ represents it provides the fixed phase shift to the $s$-th BS.
$\tilde{\mathbf{a}}_s^T \triangleq [a_{s,m}, \ldots, a_{s,M}] \in \mathbb{C}^{1 \times M}$ denotes the IRS indicator vector at the $s$-th BS, and $\mathbf{a}_m \triangleq [a_{1,m}, \ldots, a_{S,m}]^T \in \mathbb{C}^{S \times 1}$ represents the IRS indicator vector of the $m$-th element.
For example, as shown in Table \RNum{1}, for the $m$-th element with capacitance $C_1$, IRS provides a desired phase shift for the signal of 2.605 $\mathrm{GHz}$ while offers a fixed $0/2\pi$ phase shift to signals of other frequencies. Thus, at this moment the IRS indicator is $\mathbf{a}_m = [0, 0, 1]^T$.
Moreover, due to the frequency-selective characteristic of the IRS, each element can only provide desired phase shift for most one BS, i.e., $||\mathbf{a}_m||_0 \leq 1$.

Let $\bm{\theta}_s \triangleq [\theta_{1,s}, \ldots, \theta_{M,s}]^T, \forall s \in \mathcal{S}$ denote the ideal IRS phase shift vector for the $s$-th BS.
By multiplying the corresponding elements in $\bm{\theta}_s$ and $\tilde{\mathbf{a}}_s$, the phase shifts vector for the signal from $s$-th BS is restricted to reasonable values.
Thus, the practical diagonal IRS phase shift matrix for the $s$-th BS can be obtained by
\begin{equation}
        \bm{\Theta}^{\mathrm{p}}_s  \triangleq \mathrm{diag}(e^{j\bm{\theta}_s \odot \tilde{\mathbf{a}}_s}) \triangleq \mathrm{diag}(e^{j\theta^{\mathrm{p}}_{1,s}}, \ldots, e^{j\theta^{\mathrm{p}}_{M,s}}), \forall s.
\end{equation}

After propagating through the channels of both the BS-user link and the BS-IRS-user link, the signal is corrupted by additive white Gaussian noise (AWGN) $n_{k,s} \sim \mathcal{C}\mathcal{N}(0,\sigma_k^2)$. Thus, the received baseband signal at the $k_s$-th user served by the $s$-th BS can be expressed as
\begin{equation}
    y_{k_s,s}=({\mathbf{h}^{H}_{\mathrm{r},k_s,s}}
    \bm{\Theta}^{\mathrm{p}}_s
    \mathbf{G}_{s} + \mathbf{h}_{\mathrm{d},k_s,s}^H) \mathbf{w}_{k_s,s} z_{k_s,s}+n_{k_s,s}, \forall k_s, \forall s.
\end{equation}
Since the users served by different BSs have different frequencies, it is easy to eliminate the interference between each other through the receiving filter.
Thus, the SINR for the $k_s$-th user served by the $s$-th BS is given by
\begin{equation}
\gamma_{k_s,s} = \frac{|({\mathbf{h}^{H}_{\mathrm{r},k_s,s}} \bm{\Theta}^{\mathrm{p}}_s \mathbf{G}_{s} + \mathbf{h}_{\mathrm{d},k_s,s}^H)\mathbf{w}_{k_s,s}|^2}{\sum^{K_s}_{j \neq k_s}|({\mathbf{h}^{H}_{\mathrm{r},k_s,s}}\bm{\Theta}^{\mathrm{p}}_s \mathbf{G}_{s} + \mathbf{h}_{\mathrm{d},k_s,s}^H)\mathbf{w}_{j,s}|^2+\sigma_{k_s,s}^2}, \forall k_s, \forall s.
\end{equation}

\vspace{-0.3cm}
\subsection{Problem Formulation}
Our goal is to jointly optimize the transmit beamformer for all users in different BSs $\mathbf{W} \triangleq [\mathbf{W}_1,\ldots, \mathbf{W}_S]$,
the phase-shift vectors of the IRS for all BSs $\bm{\theta} \triangleq [\bm{\theta}_1, \ldots, \bm{\theta}_S]$,  and the IRS indicator matrix $\mathbf{A}$
to minimize the total transmit power for the communication system, subject to the SINR requirement of individual users.
Therefore, the joint transmit beamformer,  IRS phase shift, and IRS indicator design problem can be formulated as:
\begin{equation}
   \begin{aligned}
   \min\limits_{\bm{\theta},\mathbf{W},\mathbf{A}}  & \
   \sum^{S}_{s=1} \sum^{K_s}_{k_s=1}||\mathbf{w}_{k_s,s}||^2\\
        \textrm{s.t.}
        &\ \frac{|({\mathbf{h}^{H}_{\mathrm{r},k_s,s}} \mathbf{\Theta}^\mathrm{p}_{s}  \mathbf{G}_{s} + \mathbf{h}_{\mathrm{d},k_s,s}^H)\mathbf{w}_{k_s,s}|^2}
        {\sum^{K_s}_{j \neq k_s}|({\mathbf{h}^{H}_{\mathrm{r},k_s,s}} \mathbf{\Theta}^\mathrm{p}_{s}  \mathbf{G}_{s} + \mathbf{h}_{\mathrm{d},k_s,s}^H)\mathbf{w}_{j,s}|^2+\sigma_{k_s,s}^2} \geq \gamma_{k_s,s}, \forall k_s, s,\\
        & \ \bm{\Theta}^{\mathrm{p}}_s  = \mathrm{diag}(e^{j\bm{\theta}_s \odot \tilde{\mathbf{a}}_s}) = \mathrm{diag}(e^{j\theta^{\mathrm{p}}_{1,s}}, \ldots, e^{j\theta^{\mathrm{p}}_{M,s}}), \forall s \in \mathcal{S}, \\
        &\ ||\mathbf{a}_m||_0 \leq 1, \forall m \in \mathcal{M}.\label{eq:problem1}
    \end{aligned}
\end{equation}
where $\gamma_{k_s,s} > 0$ denotes the minimum SINR requirement of the $k_s$-th user served by the $s$-th BS.

\section{Joint Transmit Beamformer and IRS Phase Shift Design}
It is challenging to solve problem (\ref{eq:problem1}) due to the non-convex constraints in which the transmit beamforming and phase shifts are coupled.
Furthermore, due to the correlation of the IRS reflection coefficient of different BSs, it is difficult to optimize IRS phase shifts to simultaneously satisfies the requirements of different BSs.
Therefore, we first divided the multi-band problem into multiple single-band subproblems, we further define the ideal diagonal IRS phase shift matrix for the $s$-th BS $\bm{\Theta}_s = \mathrm{diag}(e^{j\bm{\theta}_s \odot \tilde{\mathbf{a}}^\mathrm{i}_s}) = \mathrm{diag}(e^{j\theta_{1,s}}, \ldots, e^{j\theta_{M,s}})$,
when the reflection co-efficient of each IRS element is shared by all BSs (i.e., $\tilde{\mathbf{a}}^\mathrm{i}_s = [1, 1, \ldots, 1]^T$).
Then the $s$-th subproblem can be expressed as
\begin{equation}
   \begin{aligned}
   \min\limits_{\bm{\Theta}_s,\mathbf{W}_s}  & \
   \sum^{K_s}_{k_s=1}||\mathbf{w}_{k_s,s}||^2\\
        \textrm{s.t.}
        &\ \frac{|({\mathbf{h}^{H}_{\mathrm{r},k_s,s}} \mathbf{\Theta}_{s}  \mathbf{G}_{s} + \mathbf{h}_{\mathrm{d},k_s,s}^H)\mathbf{w}_{k_s,s}|^2}{\sum^{K_s}_{j \neq k_s}|({\mathbf{h}^{H}_{\mathrm{r},k_s,s}} \mathbf{\Theta}_{s}  \mathbf{G}_{s} + \mathbf{h}_{\mathrm{d},k_s,s}^H)\mathbf{w}_{j,s}|^2+\sigma_{k_s,s}^2} \geq \gamma_{k_s,s}, \forall k_s,\\
        &\ |\mathbf{\Theta}_{s}(m,m)| = 1, \forall m, s.
        \label{eq:problem1d}
    \end{aligned}
\end{equation}
For each single-band problem, we propose an alternate algorithm to iteratively find the conditionally optimal solution of $\bm{\Theta}_s$ and $\mathbf{W}_s$. Then, we try to obtain the sub-optimal IRS indicator matrix $\mathbf{A}$ by linear search algorithm.


\subsection{Phase Shift and Transmit Beamforming Design for Each Single-band Problem}
Firstly, for each single-band problem, we aim to obtain the optimal phase shift matrix $\bm{\Theta}_s$ for the users served by the $s$-th BS when transmitting beamforming vector $\mathbf{W}_s$ is fixed. Since there are no variables to be optimized in the objective equation, the problem (\ref{eq:problem1d}) becomes a feasibility-check problem. In order to facilitate the alternate algorithm, we take the constraint condition in problem (\ref{eq:problem1d}) as the objective equation. By aiming to maximize the total SINR, the problem of solving $\bm{\Theta}_s$ can be expressed as

\begin{equation}
   \begin{aligned}
   \max\limits_{\bm{\Theta}_s}  &\
   \sum^{K_s}_{k_s=1} \frac{|({\mathbf{h}^{H}_{\mathrm{r},k_s,s}} \mathbf{\Theta}_{s}  \mathbf{G}_{s} + \mathbf{h}_{\mathrm{d},k_s,s}^H)\mathbf{w}_{k_s,s}|^2}{\sum^{K_s}_{j \neq k_s}|({\mathbf{h}^{H}_{\mathrm{r},k_s,s}} \mathbf{\Theta}_{s}  \mathbf{G}_{s} + \mathbf{h}_{\mathrm{d},k_s,s}^H)\mathbf{w}_{j,s}|^2+\sigma_{k_s,s}^2} \\
   \textrm{s.t.}&\ |\mathbf{\Theta}_{s}(m,m)| = 1, \forall m, s.
        \label{eq:problem2}
    \end{aligned}
\end{equation}

For this multi-cell IRS phase shift design problem, inspired by the combined channel gain maximization optimization algorithm which is commonly used in the single-user case, we obtain the phase shift at the IRS by solving a maximizing weighted effective channel gain problem. This approach aims to align with the phases of different user channels in order to maximize the beamforming gain of the IRS \cite{Two-stage}. Due to the correlative IRS phase shift affecting all users served by different BSs with different channels in problem (\ref{eq:problem2}), the combined channel power gains of different users in the different SPs cannot be maximized at the same time in general, which thus need to be balanced for optimally solving \cite{Two-stage}. This is why we weight the combined channel gain according to the SINR restrictions.
The weighted combined channel gain with transmit beamforming of users served by $s$-th BSs can be expressed as
\begin{equation}
     \sum^{K_s}_{k_s=1} \frac{1}{\gamma_{k_s,s}\sigma_{k_s,s}^2 }|{\mathbf{h}^{H}_{\mathrm{r},k_s,s}} \mathbf{\Theta}_{s}  \mathbf{G}_{s} \mathbf{w}_{k_s,s}+ \mathbf{h}_{\mathrm{d},k_s,s}^H \mathbf{w}_{k_s,s}|^2.\label{eq:combined channel}
\end{equation}

Based on (\ref{eq:combined channel}), the sub-optimal phase shift for the $s$-th BS can be obtained by solving the following problem
\begin{equation}
   \begin{aligned}
    \max_{\bm{\upsilon}_{s}} &\ \sum^{K_s}_{k_s=1}\frac{1}{\gamma_{k_s,s}\sigma_{k_s,s}^2 } |\bm{\upsilon}^H_{s}\mathbf{d}_{k_s,s} + b_{k_s,s}|^2\\
    \textrm{s.t.} &\ \ |\upsilon_{m,s}| = 1, \forall m = 1, \ldots, M,\label{eq:problem3}
   \end{aligned}
\end{equation}
where $\bm{\upsilon}_s \triangleq [e^{j\theta_{1,s}}, \ldots, e^{j\theta_{1,s}}]^H, \forall s \in \mathcal{S}$,
$\mathbf{d}_{k_s,s} \triangleq \mathrm{diag}(\mathbf{h}^{H}_{\mathrm{r},k_s,s}) \mathbf{G}_{s}  \mathbf{w}_{k_s,s}$,
and $b_{k_s,s}\triangleq \mathbf{h}^{H}_{\mathrm{d},k_s,s}\mathbf{w}_{k_s,s}, \forall k_s, s$,
denote the intermediate variables designed to facilitate subsequent optimization.

Unfortunately, the unit modulus constraint for the IRS phase shifts in (\ref{eq:problem3}) introduce the difficulty for the algorithm development due to its non-convexity. In general, non-convex relaxation and alternating minimization are used for handling this type of constraint. However, the non-convex relaxation method suffers a performance and the alternating minimization method may have slow convergence. Thus, we adopt the Riemannian-manifold-based algorithm, which is widely used in the IRS phase shift design \cite{IRS_Liuxing2} and can achieve a locally optimal solution of the original optimization problem with very fast convergence \cite{Liuxing}.

With expanding the quadratic component in (\ref{eq:problem3}) and ignoring the constant terms
$\sum_{k_s = 1}^{K_s}\frac{b_{k_s,s}^2}{\gamma_{k_s,s}\sigma_{k_s,s}^2 }$, optimization problem (\ref{eq:problem3}) can be formulated as
\begin{equation}
\begin{aligned}
\min\limits_{\mathbf{v}_{s}}\,\,\,\, &-\bm{\upsilon}_s^H\mathbf{D}_s\bm{\upsilon}_s - \bm{\upsilon}_s^H\mathbf{b}_s - \mathbf{b}_s^H\bm{\upsilon}_s\\
\textrm{s.t.} \,\,\,\, &|\upsilon_{m,s}| = 1, \forall m \in \mathcal{M},\label{eq:problem4}
\end{aligned}
\end{equation}
where we define
\begin{equation}
\begin{aligned}
    \mathbf{D}_s \triangleq & \sum_{k_s=1}^{K_s} \mathbf{d}_{k_s,s}^H \mathbf{d}_{k_s,s},\\
    \mathbf{b}_s^H \triangleq & \sum_{k_s=1}^{K_s} b_{k_s,s}\mathbf{d}_{k_s,s}.
\end{aligned}
\end{equation}
It can be easily obtained that
derivative of the target equation in problem  (\ref{eq:problem4}) is $-2\mathbf{D}\bm{\upsilon}_s - 2\mathbf{b}_s^H.$
Thus, we can acquire the solution by \textit{Riemannian manifold} conveniently \cite{Liuxing}. Due to the space limitation, the derivation of \textit{Riemannian manifolds} is omitted in this paper.

After acquiring the IRS phase shift matrix $\bm{\Theta}_{s}$, we need to calculate the transmit beamforming at the $s$-th BS.
For given IRS phase shift $\bm{\Theta}_{s}$, the combined channel from the $s$-th BS to the user $k_s$ can be expressed as $\mathbf{h}_{k_s,s}^H \triangleq {\mathbf{h}^{H}_{\mathrm{r},k_s,s}} \mathbf{\Theta}_{s}  \mathbf{G}_{s} + \mathbf{h}_{\mathrm{d},k_s,s}^H$.
Thus, problem (\ref{eq:problem1}) is reformulated to
\begin{equation}
   \begin{aligned}
    \min\limits_{\mathbf{W}_s}  &\ \
    \sum^{K_s}_{k=1}||\mathbf{w}_{k_s,s}||^2\\
        \textrm{s.t.}
        &\ \ \frac{|\mathbf{h}_{k_s,s}^H\mathbf{w}_{k_s,s}|^2}{\sum^{K_s}_{j \neq k_s}|\mathbf{h}_{k_s,s}^H\mathbf{w}_{j,s}|^2+\sigma_{k_s,s}^2} \geq \gamma_{k_s,s}, \forall k_s. \label{eq:problem5}\\
    \end{aligned}
\end{equation}
In order to solve the problem (\ref{eq:problem5}) easily, we reformulate this non-convex constraint as
\begin{equation}
   \begin{aligned}
    \min\limits_{\mathbf{W}_s}  &\ \
    \sum^{K_s}_{k_s=1}||\mathbf{w}_{k_s,s}||^2\\
        \textrm{s.t.}
        &\ \ \sqrt{1+\gamma_{k,s}}{|\mathbf{h}_{k_s,s}^H\mathbf{w}_{k_s,s}|}-
        \sqrt{|\mathbf{h}_{k_s,s}^H\mathbf{W}_{s}|^2+\sigma_{k_s,s}^2}
        \geq 0, \forall k_s \label{eq:problem6},
    \end{aligned}
\end{equation}
which is a typical second-order cone programming (SOCP) problem and hence can be easily solved by the popular convex optimization solvers such as CVX \cite{CVX}.
Now, the complete procedure can be established by iteratively optimizing $\mathbf{\Theta}_s^\star$ and $\mathbf{W}_s^\star$ until the convergence is found.

\begin{algorithm}[t]
\caption{Joint transmit beamformer and IRS phase shift design}
\label{alg:SH}
    \begin{algorithmic}[1]
    \REQUIRE $\mathbf{h}_{\mathrm{r},k_s,s}^H , \mathbf{G}_s, \mathbf{h}_{\mathrm{d},k_s,s}^H , \sigma_{k_s,s}, \gamma_{k_s,s}$.
    \ENSURE $\bm{\theta}^{\star}, \mathbf{W}^{\star}, \mathbf{A}^\star$.
    \STATE {Initialize $ \bm{\theta_{s}} = [\theta_{1,s}, \ldots, \theta_{M,s}]^T, s = 1, \ldots, S$. }
    \FOR {$s =1$ to $S$}
        \STATE {Calculate $\mathbf{W}_{s}^{\star}$ by solving problem (\ref{eq:problem6}) with CVX.}
        \STATE {Calculate $\bm{\Theta}_{s}^{\star}$ by solving problem (\ref{eq:problem4}) with \textit{Riemannian manifold}}.
        \STATE {Goto step 2 while no convergence of $\sum^{K_s}_{k=1}||\mathbf{w}_{k,s}||^2$. }
    \ENDFOR
        \FOR {$m =1$ to $M$}
            \FOR {$s =1$ to $S+1$}
                \STATE Assign a value to $\mathbf{a}_m.$
            \ENDFOR
            \STATE {Find the optimal IRS indicator vector $\mathbf{a}_m^\star$ for the $m$-th element, which has the minimum total transmit power.}
       \ENDFOR
        \STATE {Calculate $\mathbf{W}^{\star}$ by solving problem (\ref{eq:problem6}) with CVX.}
    \end{algorithmic}
\end{algorithm}

\subsection{Service Provider Selection of IRS}
Though the above alternate algorithm, we can obtain the optimal phase shift solutions $\bm{\Theta}^\star_s$ and transmit beamformer $\mathbf{W}^\star_s$ for each BS. So far, this problem can transform into an optimal BS selection problem, and we turn to search the optimal IRS indicator matrix $\mathbf{A}$.

To obtain the optimal IRS indicator $\mathbf{A}$, we should search for $s$ possible situations of each $\mathbf{a}_m$. This exhaustive search involves trying as many as $s^m$ possible situations, which is unaffordable. Thus, when we search for the BS served by the $m$-th element, we fix the service states of other elements to significantly reduce the algorithm complexity, then this problem can be expressed as
\begin{equation}
   \begin{aligned}
   \min\limits_{\mathbf{A}}  & \
   \sum^{S}_{s=1} \sum^{K_s}_{k_s=1}||\mathbf{w}_{k_s,s}||^2\\
        \textrm{s.t.}
        &\ \frac{|({\mathbf{h}^{H}_{\mathrm{r},k_s,s}} \mathbf{\Theta}^\mathrm{p}_{s}  \mathbf{G}_{s} + \mathbf{h}_{\mathrm{d},k_s,s}^H)\mathbf{w}_{k_s,s}|^2}
        {\sum^{K_s}_{j \neq k_s}|({\mathbf{h}^{H}_{\mathrm{r},k_s,s}} \mathbf{\Theta}^\mathrm{p}_{s}  \mathbf{G}_{s} + \mathbf{h}_{\mathrm{d},k_s,s}^H)\mathbf{w}_{j,s}|^2+\sigma_{k_s,s}^2} \geq \gamma_{k_s,s}, \forall k_s, s,\\
        & \ \bm{\Theta}^{\mathrm{p}}_s  = \mathrm{diag}(e^{j\bm{\theta}_s \odot \tilde{\mathbf{a}}_s}) = \mathrm{diag}(e^{j\theta^{\mathrm{p}}_{1,s}}, \ldots, e^{j\theta^{\mathrm{p}}_{M,s}}), \\
        &\ ||\mathbf{a}_m||_0 \leq 1, \forall m \in \mathcal{M}.\label{eq:problem7}
    \end{aligned}
\end{equation}
After having the optimal IRS indicator matrix $\mathbf{A}^\star$, we should obtain the optimal transmit beamformer $\mathbf{W}^\star$ by solving problem (\ref{eq:problem6}) with CVX. The proposed IRS reflect beamforming design is summarized in Algorithm 1.

\section{Simulation Results}
We consider a multi-cell multi-band system consisting of $S =3$ BSs with $N_t = 16$ antennas, $K_s = 4, \forall s$, single-antenna users served by each BS, and an IRS with $M $ reflecting elements to aid the communication.
Both IRS and users are located 50 meters (m) apart from BS.
Moreover, the user is randomly located around the IRS with 2m distance between them.
We assumed that all the channels involve Rayleigh fading and the signal attenuation at a reference distance of 1m is set as 30 dB.
The path loss exponents are set to 2.5, 2.8, and 3.5 for the channels between BS-IRS, IRS-user, and BS-user, respectively.
\begin{figure}[t]
\centering
  \includegraphics[height = 3  in]{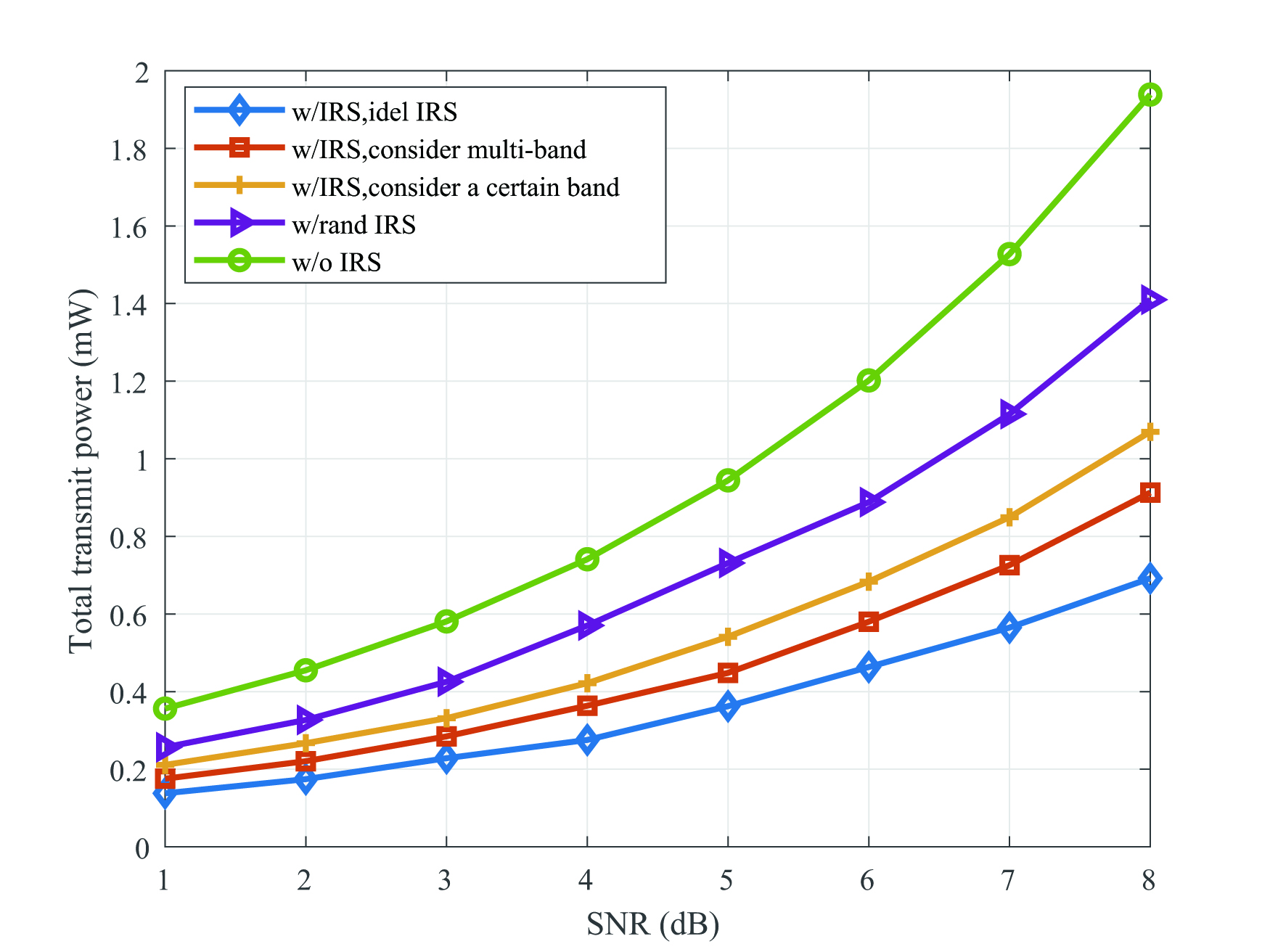}
  \caption{Total transmit power versus SINR ($M = 64, N_t = 16, S =3$).}\label{fig:Result}
  \vspace{-0.4 cm}
\end{figure}

\begin{figure}[t]
\centering
  \includegraphics[height = 3 in]{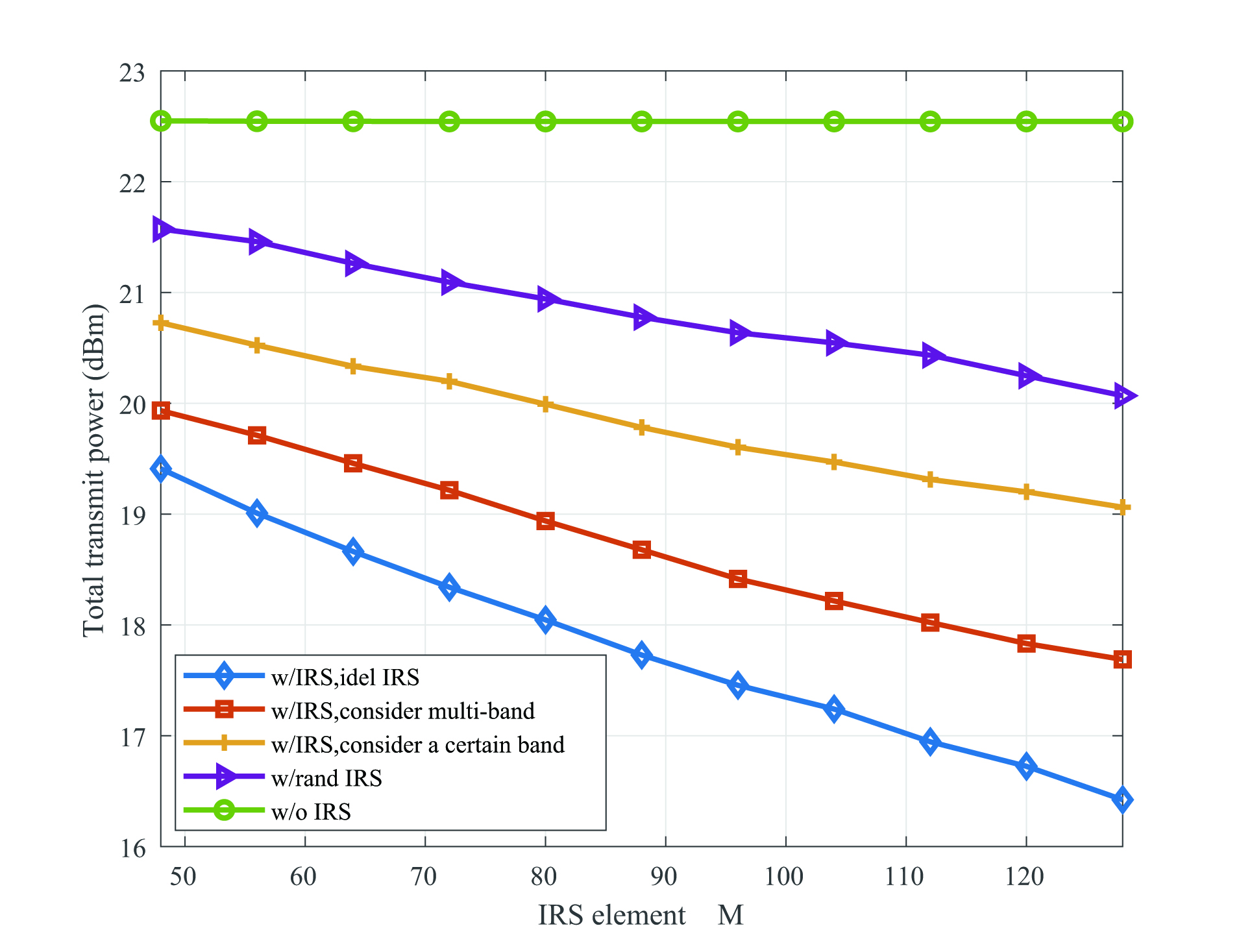}
  \caption{Total transmit power versus the number of IRS elements $M$ ($\gamma_{k_s,s} = 5\mathrm{dB}, N_t = 16, S =3$).}\label{fig:Result2}
  \vspace{-0.7 cm}
\end{figure}

Fig. \ref{fig:Result} shows the total transmit power by considering the influence due to the multi-band for optimizing the IRS system.
In the simulation results, for comparison, the curve with the legend ``w/IRS and ideal model'' is used to be an upper band and shows the total transmit power achieved when employing ideal IRS which can provide optimal phase shift for signals with different frequencies during beamforming design and compute performance
The curve with the legend ``w/IRS consider multi-band'' shows the total transmit power achieved when considering the influence due to the multi-band during beamforming design and compute performance, i.e., a practical IRS which can provide restricted phase shift for different users is deployed to assist the communication system, and the proposed algorithm is applied to design the IRS user selection.
The curve with the legend ``w/IRS consider a certain band'' shows the total transmit power achieved when considering the influence due to the multi-band but without applying the IRS user selection algorithm, i.e., the practical IRS only serves the users in a certain band, but do not provide desired phase shift for users in other bands.
In addition, the performance with random IRS (w/rand IRS, ideal model) and without using the IRS (w/o IRS) is also included as a benchmark.
It is observed from Fig. \ref{fig:Result} that, while the IRS can significantly improve performance, our proposed practical model and user selection algorithm can provide remarkable better performance than the ideal model without user selection for optimizing the IRS-aided multi-frequency communication system.
Due to the frequency selectivity of the IRS, the more users with different frequencies, the more phase-frequency distortion will be produced, which causes difficulty on the reflect beamforming optimization.
Moreover, Fig. \ref{fig:Result2} illustrates the performance of different schemes as a function of the number of IRS elements.
A similar conclusion can be drawn as in Fig. \ref{fig:Result}.
\section{Conclusions}
In this paper, we proposed an intelligent reflecting surface (IRS)-assisted multi-cell multi-band system, with base stations (BSs) operating on different frequencies and consider the influence between signals with different frequencies based on the IRS practical phase shift model.
Based on this realistic communication scenario, joint transmit beamformer and IRS phase shift optimization algorithm was introduced for an IRS-aided multi-cell multi-band  communication system.
Simulation results demonstrated the significant performance improvement by utilizing the proposed practical model in optimizing the IRS-aided multi-cell multi-band  system.
While in this initial work a heuristic IRS design algorithm was developed to illustrate the importance of considering the frequency-dependent IRS  in the multi-cell multi-band scenario, possible directions for the future studies include expanding multi-IRS allocation scenario, multi-cell game scenario and the design of more efficient algorithms.

\end{document}